\documentclass[12pt,preprint]{aastex}

\def\ltsima{$\; \buildrel < \over \sim \;$}
\def\simlt{\lower.5ex\hbox{\ltsima}}

\newcommand{\dgr}{{\hbox {$^\circ$}}}
\newcommand{\pL}{$p_{L}$}
\newcommand{\pC}{$p_{C}$}

\shorttitle{XPOL at the IRAM 30m telescope}
\shortauthors{Thum et al.}

\slugcomment{\it Accepted for publication by Publications of
    the Astronomical Society of the Pacific on 20/05/2008}

\begin{document}

\title{XPOL --- the correlation polarimeter at the IRAM 30m telescope}

\author{C. Thum, H. Wiesemeyer\altaffilmark{1}}
\affil{Institut de Radio Astronomie Millim\'etrique,
    Domaine Universitaire de Grenoble, 300 Rue de la Piscine, 
    38406 St.~Martin--d'H\`eres, France}
\author{G. Paubert and S. Navarro}
\affil{Instituto de Radio Astronom{\'\i}a Milim\'etrica, N\'ucleo Central, 
       Avd. Divina Pastora No. 7--9, 18000 Granada, Spain}
\and
\author{D. Morris}
\affil{Institut de Radio Astronomie Millim\'etrique,
    Domaine Universitaire de Grenoble, 300 Rue de la Piscine, 
    38406 St.~Martin--d'H\`eres, France}

\altaffiltext{1}{present address: Instituto de Radio Astronom{\'\i}a
       Milim\'etrica, N\'ucleo Central,  
       Avd. Divina Pastora No. 7--9, 18000 Granada, Spain  }


\begin{abstract}
XPOL, the first correlation polarimeter at a large millimeter telescope,
uses a flexible digital correlator to measure all four Stokes
parameters simultaneously, 
i.e. the total power $I$, the linear
polarization components $Q$ and $U$, and the circular polarization
$V$. The versatility of the backend provides  
adequate bandwidth for efficient  continuum observations as well as
sufficient spectral resolution (40 kHz) for observations of narrow lines. We
demonstrate that the polarimetry specific calibrations are handled
with sufficient precision, in particular the relative phase between the
Observatory's two orthogonally linearly polarized receivers. 

The many facets of instrumental polarization are studied at 3mm
wavelength in all Stokes parameters: on--axis with point sources and
off--axis with beam maps. Stokes $Q$ which is measured as the power
difference between the receivers is affected by instrumental
polarization at the 1.5\% level. Stokes $U$ and $V$ which are measured as 
cross correlations are very little affected (maximum sidelobes 
0.6\% ($U$) and 0.3\% ($V$)). These levels critically depend on the
precision of the receiver alignment. They reach these minimum levels
set by small ellipticities of the feed horns when alignment is optimum
($\simlt0.3''$).  A second critical prerequisite for low polarization
sidelobes turned out to be the correct orientation of the polarization
splitter grid. Its cross polarization properties are modeled in detail. 

XPOL observations are therefore limited only by receiver noise in
Stokes $U$ and $V$ even for extended sources. Systematic effects set
in at the 1.5\% level in observations of Stokes $Q$. With proper
precautions, this limitation can be overcome for point sources. Stokes
$Q$ observations of extended sources are the most difficult with XPOL.
\end{abstract}



\section{Introduction}

Spectropolarimetric observations at millimeter wavelengths are of
considerable astrophysical interest. Several molecular transitions
which are bright in star forming regions are polarized due to the
magnetic fields associated with most stages of star formation
\citep{Vallee}. Lines may either be circularly polarized due to the
Zeeman effect or linearly polarized through the Goldreich--Kylafis
effect, or both. Line polarization is also known in late stages of
stellar evolution where  dense circumstellar shells often give rise to
maser emission in millimeter lines \citep{Menten}. At least in the case of SiO,
the transitions may have very complex polarization characteristics due
to magnetic fields and maser propagation. 

The millimeter continuum may also be polarized in Galactic sources
due to emission from non-spherical paramagnetic dust grains.  The
continuum radiation of extragalactic sources, notably the synchrotron
emission from active galactic nuclei, is also often polarized.  Its
detection at millimeter wavelengths is of interest, since the lower
optical depth allows us to look deeper into the sources than is
possible at longer wavelengths and since these wavelengths are less
affected by interstellar scattering.  

The ideal polarimeter should therefore be capable of observing lines
as well as continuum, and all four Stokes parameters should be
measured. Given the usual weakness of the polarization signal in the
millimeter range, a large telescope is advantageous if compact sources
are to be measured.  The few spectral polarization observations made previously
at the IRAM 30m telescope measured  
circular polarization \citep{Dick1,Dick2,TM99}. Linear polarization
measurements were attempted in the continuum with a bolometer 
\citep{Lemke}.    
Many more observations of linear polarization were made at the
James--Clark--Maxwell telescope \citep[see][and references therein]{Jane}.

In October 1999 when we began developing XPOL, the IRAM 30m
telescope had started a profound upgrade of its receivers.  At the end of
this transformation, the telescope was equipped with 4 single beam, dual
polarization heterodyne  receivers which cover most of the 75 to 270
GHz frequency range. All
receivers have linearly polarized feed horns. Here we concentrate on
work at 3mm wavelength with receivers A\,100 (vertically polarized) and
B100 (horizontally polarized). The four Stokes parameters  are derived
from the IF signals from these receivers (the Intermediate Frequecy is
at 1.5 GHz) and refer to the
right handed Nasmyth coordinate system $\mathcal{K_N}$
(Fig.~\ref{f:schema}) in which these receivers are stationary. 
In this configuration, the Stokes parameters relate to the time--averaged
horizontal, $E_x$, and vertical, $E_y$, electric fields and their 
relative phase $\delta$  \citep[chap.~10]{BW}:
\begin{eqnarray}
I  =   \langle E_x^2\rangle + \langle E_y^2\rangle \label{e:Stokes} \\
Q  =   \langle E_x^2\rangle - \langle E_y^2\rangle  \\
U  =   2\ \langle E_x E_y\,{\rm cos}\,\delta\rangle   \\
V  =   2\ \langle E_x E_y\,{\rm sin}\,\delta\rangle   
\end{eqnarray}
This description of the Stokes parameters in terms of electric fields
lends itself most naturally to our case where heterodyne receivers
amplify the incoming electromagnetic fields.
Stokes $I$ and $Q$ are thus obtained from power measurements, whereas
Stokes $U$ and $V$ are measured as correlations. 
The fractional linear, \pL, and circular, \pC, polarization and the
polarization angle, $\tau$, are then derived in the usual way.
  
Correlation polarimeters are well known in radio astronomy
\citep[chapter 4.7]{RW}. Heiles and collaborators
\citep{Heiles-I,Heiles-All,Heiles-Mueller} give a very detailed
description of this measurement technique as used at the Arecibo
Observatory. To our 
knowledge, XPOL at the IRAM 30m telescope and its analog predecessor IFPOL 
\citep{IFpol} are the first such correlation polarimeters
 used at a large millimeter telescope.

\section{Instumental setup}

The procedure which we designated XPOL, of cross correlation
spectro\-polarimetry at the IRAM 30m telescope, makes use of existing
observatory 
equipment as much as possible. Notably, we use the two single beam 3mm SIS
receivers which are housed in separate dewars. This section describes
the standard equipment inasmuch as relevant for polarization
observations and the modifications made for XPOL.

\subsection{Nasmyth optics}
\begin{figure}
\epsscale{.71}
\plotone{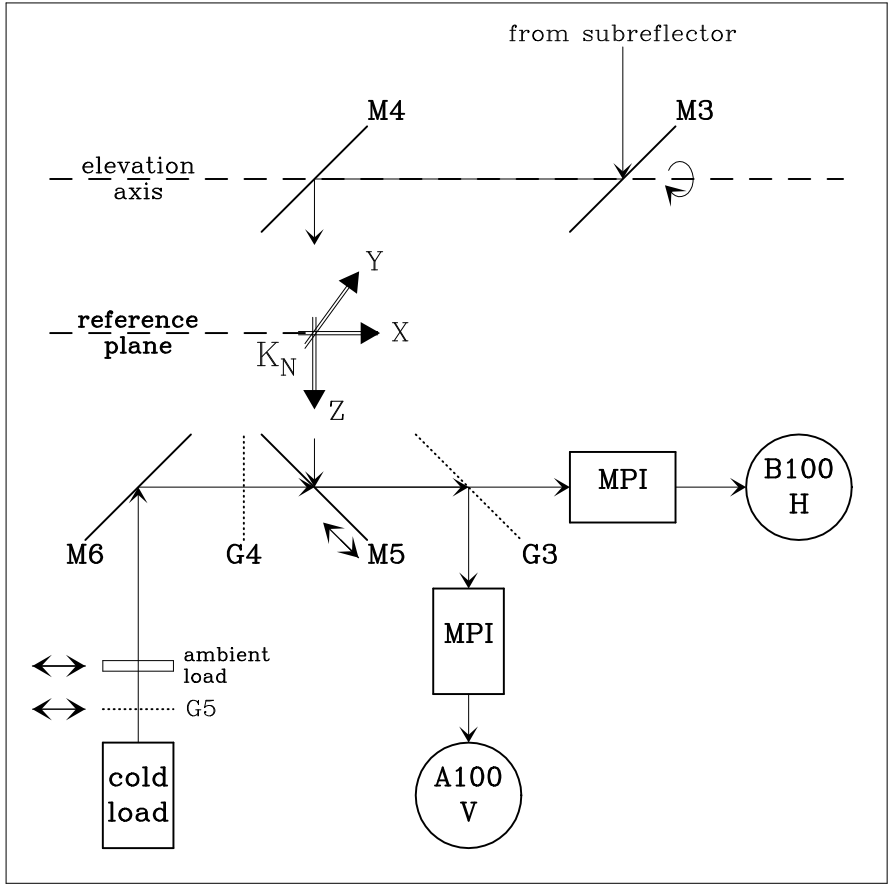}
\caption{Schematic of the optical paths in the Nasmyth cabin of the
  30m telescope. Radiation coming down from the subreflector is
  relayed by several mirrors, of which M3 rotates about the elevation
  axis, to the wire grid G3 which reflects vertically linearly
  polarized power to the 3mm receiver A100 and transmits horizontal power
  to B100. The first beam waist is located at the Martin--Puplett
  interferometers, MPI. When mirror M5 is retracted from the beam, as
  indicated by the $\Longleftrightarrow$ symbol,
  both receivers look at mirror M6 which focuses their beams onto the
  calibration unit. Wire grid G4 is occasionally inserted for
  calibration  of  the orientation of G5 (see sect.~\ref{sss:method}).
  We define the polarization parameters in the
  righthanded coordinate system $\mathcal{K_N}$, stationary in the 
  Nasmyth cabin and
  located at the reference plane where $X$ is transverse  horizontal, 
  $Y$ is vertical (pointing upward), and $Z$ points in the direction
  of beam propagation. 
\label{f:schema}
}
\end{figure}

The 3mm receivers A\,100 and B\,100 located in the Nasmyth cabin view
the subreflector through a series of optical elements shown very
schematically in Fig.~\ref{f:schema}. We refer all polarization
measurements to the coordinate system $\mathcal{K_N}$ stationary in
the Nasmyth cabin as shown in Fig.~\ref{f:schema}. Mirrors M3, M4, and
M5 bring the incoming beam to the wire grid G3 which reflects vertical 
(parallel to the $Y$--axis of $\mathcal{K_N}$) polarization to
A\,100 and transmits horizontal (parallel to $X$) polarization to B\,100.  
In the default setting, the wires of G3 are parallel to the plane of
incidence. The Martin--Puplett interferometers (MPI) can be used to rotate the
polarization direction for matching with the respective orientations
of the 3mm and 1.3mm receivers (housed in the same dewars, but not
shown here). The MPIs are located at the first beam waists. Additional
optics, not shown, refocus the beams onto the receiver feed horns.

\placefigure{f:schema}

When mirror M5 is retracted from the beam, the receivers view M6 which
focuses the beams onto the calibration unit which is equipped with the
usual ambient and cold loads for the calibration of antenna
temperature. A wire grid, G5, is employed for the calibration of the
relative phase $\delta$ (sect.~\ref{ss:phasecal}).

\subsection{Modifications for XPOL}
\label{ss:XPOL}
The standard instrumental setup was modified in several respects for
the polarization measurements. First, the local oscillator signal
is normally derived from separate synthesizers (operating near 5 GHz and 
100 MHz) for each receiver. In polarimetry, the two receivers share
both the 5 GHz and the 100 MHz synthesizers.
In this way, the phase noise between the two receivers is reduced to
an acceptable level (see sect.~\ref{ss:calUV}).

The actual correlation between the two receiver signals is made in the
spectral backend VESPA (VErsatile Spectroscopic and Polarimetric
Analyzer) which is located in the observatory building
and connected to the receivers in the Nasmyth cabin by circa 100m long
coaxial cables. The length difference of these cables, originally found to be
about 0.3m, gave rise to a very steep phase variation across the 500
MHz bandpass, the maximum bandpass available with the 3mm receivers.
We adjusted the length of these cables until the global phase gradient
was smaller than phase structure introduced by  other components
(Fig.~\ref{f:phasecal}). The residual difference of the 
electric paths is smaller than 5 cm. The detailed spectral behavior
of the phase and its calibration is described below (sect.~\ref{ss:phasecal}).

In non--polarimetric observations, the Martin--Puplett
interferometers, MPI in Fig.~\ref{f:schema}, are tuned to a path
difference which permits simultaneous maximum transmission of the 3
and 1.3mm frequencies observed\footnote{Receivers at 1.3mm, A230 and
  B230, are housed in the same dewars as A100 and B100, but are not
  shown on Fig.~\ref{f:schema}.}.
The compromise setting implies a very small
loss at each frequency. With XPOL  we do not normally use the
1.3mm receivers in parallel with the 3mm receivers;  although this
operation mode is feasible, we tune the MPIs such that only the 3mm
transmission is maximized, so that  no losses occur.

The most profound modification was required for the spectral backend,
which had to be capable of making cross correlations in addition to the
autocorrelations, which are normally all that is needed  on a
single dish telescope.  This substantial upgrade became practical
after the Plateau de Bure (cross)correlator  became available
following the addition of a new antenna and new correlator to the IRAM
interferometer. 
The integration of these components with the previous 30m correlator
created a larger bandwidth correlator, dubbed VESPA \citep{GP}. It is  
capable of simultaneously making the auto-- and cross correlations
required for the simultaneous  measurement of the four Stokes parameters
(eqns.\,1 -- 4). The data
streams are organized such that auto-- and cross correlations of each
spectral subband are derived from the same digitizer, thus sharing the entire
analog signal path.
This greatly
facilitates calibration (sect.~\ref{ss:phasecal}). VESPA is an XF--type
correlator which calculates the auto-- and complex cross correlation
functions 
and accumulates them during one observation phase (typically 2 sec in
wobbler--switching), before their Fourier transforms are written on disk.

The VESPA hardware permits the following resolution/bandwidth combinations
(in kHz/MHz) in polarimetry mode: 40/120,
80/240, 625/480\footnote{ A further special mode is available with spectral
  resolution of 2.5 MHz and a bandpass of 960 MHz split into
  two 480 MHz subbands of identical IF coverage.}.
The latter combination is the current optimum for
continuum observations where the spectral channels are averaged to get
one single power measurement for each Stokes parameter. The higher
spectral resolution modes are used for line observations where the
available bandwidth can, within limitations, be split into several
sections at separate IF frequencies.

\section{Calibration}

Polarization observations as described above invoke the precise
measurement of the four Stokes powers (eqns.\,1--4) rather than just one
measurement as in the usual total power observations. 
The Stokes $I$ and $Q$ powers are
immediately derived from the vertically and horizontally polarized
receivers A\,100 and B100 which are calibrated using the traditional
hot/cold  load technique (see, e.g. \citet[]{DD}). Calibration of the $U$
and $V$ powers invoke new issues discussed in this section: the
measurement of the relative phase between the receivers  
(sect.~\ref{ss:phasecal}), decorrelation losses (sect.~\ref{ss:calUV}),
and the signs of the Stokes parameters (sect.~\ref{ss:calPol}).
 
\subsection{Calibration of phase}
\label{ss:phasecal}

The phase $\delta$ between the receivers A\,100 and B\,100 is derived
for each spectral channel from the real 
and imaginary part of the cross correlation measured with 
VESPA.  The intrinsic phase between orthogonally linearly polarized
components of the incoming radiation field may be modified in many
ways as the radiation passes down the signal chain.
The most notable mechanisms potentially 
affecting $\delta$ are {\it (i)} differences in the geometric paths through
telescope and cabin optics up to the mixers, {\it (ii)} conversion of the
electric field into voltages at the feed horns, {\it (iii)} amplification,
filtering and transport of the signals downconverted to the first IF,
{\it (iv)} transport in the long IF cables to the backend, and 
{\it  (v)}  downconversion to basebands in VESPA with the associated
filtering and amplification. Stages  {\it (i)} and  {\it (ii)},  where
the signals are at the radio frequency, are a priori the most likely
ones to introduce differential phase shifts.  A phase calibration unit
is therefore 
best placed high up in the signal chain where it calibrates a maximum
of these phase--affecting components.

\subsubsection{Method}
\label{sss:method}
Our method of phase calibration makes use of the hot/cold calibration unit
located in the Nasmyth cabin (Fig.~\ref{f:schema}). Any phase shifts
occuring at stage {\it (i)} in the complicated optical path between
the receivers and M5 are therefore calibrated. Phase shifts picked up
in front of M5 are however not calibrated. They manifest themselves as
a component of the instrumental polarization which varies with elevation
(sect.~\ref{ss:ge}). 

The phase calibration of XPOL consists of the 3 subscans of the
standard temperature calibration (on sky, ambient 
load, and cold load) and an additional observation of the cold load 
(subscan No.\,4) where the wire grid G5 is introduced into the beam
(Fig.\ref{f:schema}). The 
unpolarized hot--cold signal is thus converted into a strong linearly
polarized signal which covers the entire bandwidth of the backend. The
angle of the wires is fixed to 53\degr, which is close enough to
45\degr\ to transmit comparable power to both receivers, but distinct enough
from 45\degr\ to avoid 90\degr\ ambiguities. The correct setting of this
important angle was verified by temporarily introducing another wire 
grid G4 (Fig~\ref{f:schema}). At this location, the angle which the
wires make with the horizontal plane can be precisely measured.
This angle is $54\pm0.5$\degr. Its uncertainty represents the ultimate
limit to the precision of XPOL polarization angle measurements.

\begin{figure}
\resizebox{\textwidth}{!}{\includegraphics[angle=270]{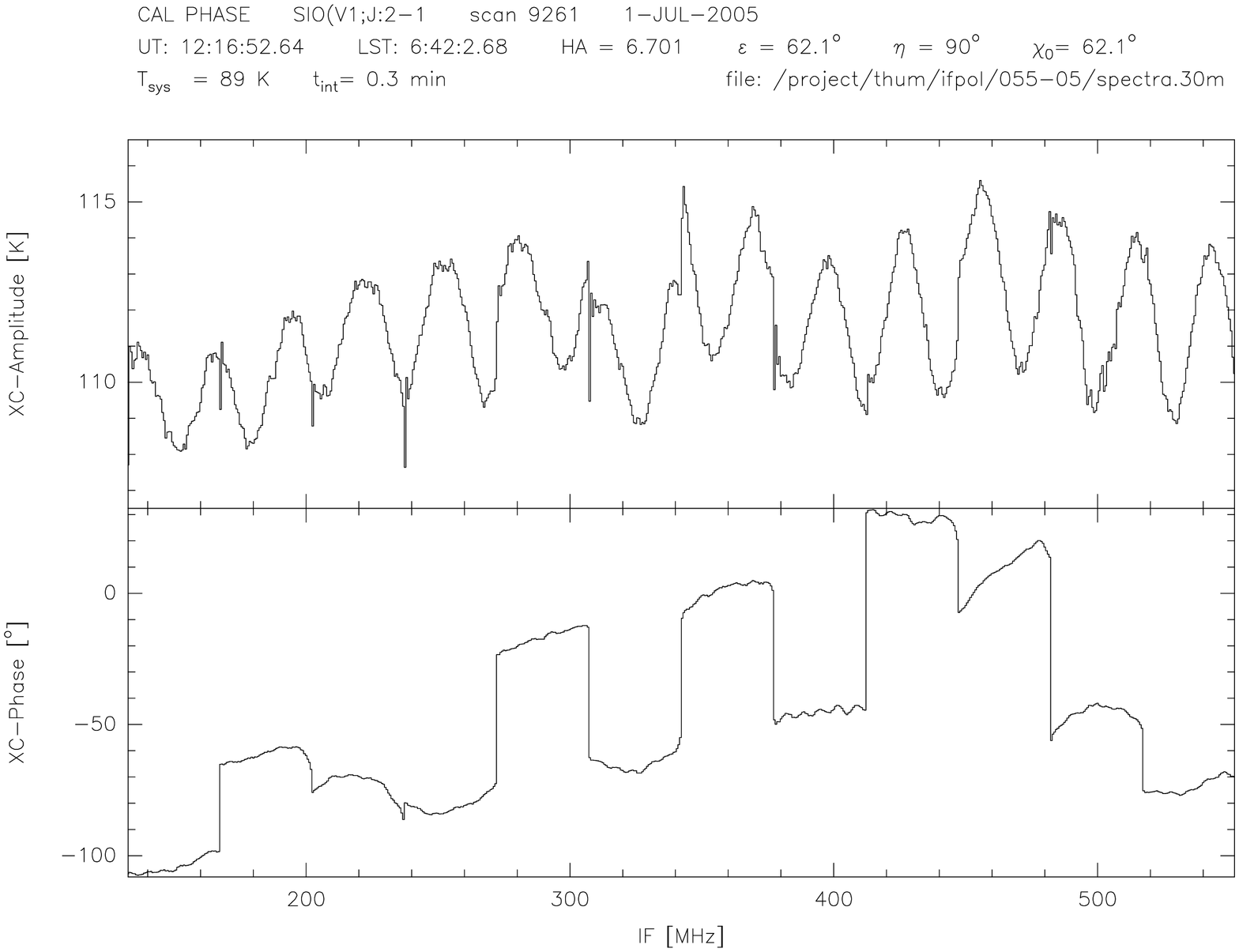}
}
\caption{Phase calibration with VESPA in broadband mode. The 480 MHz
  global bandpass is split up into 12 basebands which have distinctly
  individual bandpass characteristics. For each of the 765 spectral
  channels the amplitude ({\it upper panel}) and phase ({\it lower panel})
  is derived with precisions of $\le1$\% and $\le1$\degr, respectively,
  except for a few channels at subband  edges. The inconsequential
  standing wave of period 29.1 MHz in the amplitude spectrum 
  (see sect.~\ref{sss:precision}) arises from reflections between the 
  receiver and the calibration   unit separated by 5.1 m.
\label{f:phasecal}
}
\end{figure}

\placefigure{f:phasecal}

>From the 4 subscans of a phase calibration measurement 
the XPOL software calibrates the measured complex
cross correlation spectrum in terms of antenna temperature. 
Fig.~\ref{f:phasecal} shows amplitude and phase of such a phase
calibration. VESPA is set to its broadest bandwidth (480 MHz) where calibration
is most critical. In this mode, the bandpass is made up of 12 
basebands generated by 12 image rejection mixers which are pumped by
their respective L03 local oscillators. The phase spectrum clearly
shows this subband structure whereas the amplitude spectrum is smooth
 after calibration. The residual amplitude ripple which is due to
reflections between the receivers and the cold load is of low relative
amplitude and has no practical consequences (see sect.~\ref{sss:precision}).

\subsubsection{Precision}
\label{sss:precision}
The phase spectrum consists of 3 components: an overall gradient
across the bandpass, systematic offsets of each baseband, and an irregular
phase bandpass for each baseband. The offsets are due to differences
in the cable length of each of the twelve LO3. These differences could be
made small or negligible, but they are inconsequential, since they can
be precisely calibrated and they are stable in time due to the
location of VESPA in an air--conditioned room. The same holds
for the detailed shape of the phase bandpasses which are related to the
individual filters and amplifiers for each baseband.  No effort was
therefore made to adjust their characteristics, since the
signal--to--noise ratio  available is clearly sufficient to allow
rapid calibration of these effects on a channel--by--channel basis. 

Line observations often use smaller bandwidths where fewer and
different VESPA subbands are employed. Our phase calibration method
works correctly for all VESPA polarimetry configurations
(sect.~\ref{ss:XPOL}). In particular, since the phase of each channel
is measured, no assumptions are needed for the behavior of the phase
across the receiver bandpass. 

The small amplitude ripple ($\sim1$\%) of the phase calibration spectrum
(Fig.~\ref{f:phasecal}) introduces a sinusoidal phase error of
peak--peak amplitude $\sim0.6$\degr. This is a very small error which
furthermore tends to cancel out in broadband (continuum)
observations. The error might be detectable in spectral observations
where the line is strongly linearly polarized. As this error exchanges
power between Stokes  $U$ and $V$, it generates instrumental $V$ at
the level of 0.01$\cdot U$  at the spectral channels
near the maximum of the ripple amplitude and smaller elsewhere. The
error is therefore negligible.

The overall gradient of the phase across the 500 MHz bandpass,
due to a small residual length difference of the IF cables, is 
of the order of the baseband offsets. This phase pattern is very
stable, the (bandpass averaged) mean phase not usually varying by more
than 10\degr\ during a night. Irregular phase jumps, like those 
expected from cable stress in the azimuth cable spiral or in the large 
elevation cable wrap, are not observed. 

The only occasion when small phase jumps are seen is after 
the observing frequency has changed, e.g. when a new target source is
observed. These phase changes are proportional to the frequency
changes, and may be explained by small non-linearities in the frequency 
multiplication chain of the local oscillators.    
Phase variations due to changes in the telescope structure
are small and have not been clearly detected (sect.~\ref{ss:ge}).

\subsection{Calibration of Stokes $\mathbf U$ and $\mathbf V$ spectra}
\label{ss:calUV}

Stokes $U$ and $V$ spectra are obtained from cross correlation of the two
receivers (eqns.\,3--4). The correlated power may therefore be
affected  by phase 
noise which is a well known issue for interferometers, but not normally
of concern at single dish telescopes. If uncorrected, these losses
introduce a systematic bias into the  measurements of the linear and
circular fractional polarization degrees and the polarization angle.
We have used two methods to measure this loss which is proportional
to the observing frequency. The first one uses a strongly
polarized celestial source and is very time consuming. The second
method exploits our new scheme of calibrating the relative phase. It
therefore does not cost any extra observing time and its result is
very precise. It is now routinely used by the XPOL calibration software.

\subsubsection{Celestial source}
\label{sss:TX}
The field of view of a receiver in the Nasmyth focus is subjected to
two rotations: {\it (i)} the parallactic rotation due to the
alt-azimuth mount of the telescope, and {\it (ii)} the Nasmyth rotation
about the elevation axis. Stokes $Q$ and $U$ of a linearly polarized
celestial source therefore vary sinusoidally as the source is
tracked. The amplitudes 
of the normalized sine waves, $Q/I$ and $U/I$, must be the same and
equal to the fractional linear polarization \pL. Decorrelation
losses, however, make $U$ smaller than $Q$, an easily observable
effect on a strongly linearly polarized source.

\begin{figure}
\epsscale{.80}
\resizebox{0.70\textwidth}{!}{\includegraphics[angle=270]{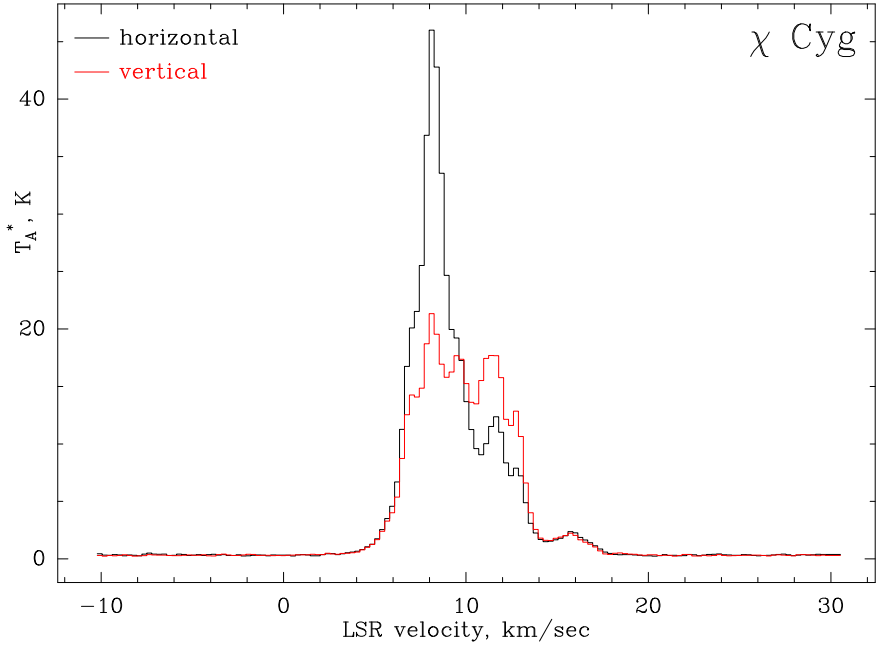}}
\resizebox{0.60\textwidth}{0.7\textwidth}{\includegraphics{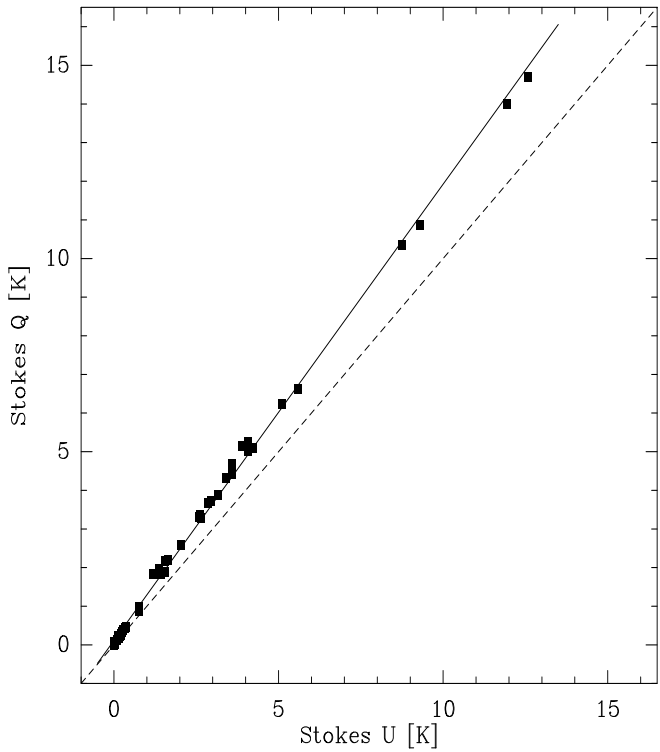}}
\caption{Measurement of the correlation loss using the strongly
  polarized  86.243 GHz SiO emission from the star $\chi$\,Cyg.
  {\it top:} 58 observations like the one shown here were made, during
  which the Nasmyth polarization angle $\tau$ 
  rotated by more  than one period. {\it bottom:} The resulting amplitude, as
  measured in the Nasmuth system $\mathcal K_N$, of the $U$
  sine wave is smaller  than that of $Q$ for all channels as indicated
  by the departure from the $45$\dgr\ line (dashed). The systematic
  difference  is ascribed to decorrelation  losses in $U$.   
\label{f:TX}
}
\end{figure}

\placefigure{f:TX}

We oberved the SiO maser star $\chi$ Cyg at 86.243 GHz (the $v=1$,
J$=2\rightarrow1$ transition) during 9 hours. The star's position
(declination $\sim33$\degr) makes it culminate near the zenith, 
where more than  a full period of the $Q$ and $U$ rotation is readily
observed. The spectrum 
has several features which are linearly polarized, as seen from the large    
difference in the spectra from the orthogonally polarized receivers
(Fig.~\ref{f:TX}, top). We have obtained 58 such spectra per observing
period from
which we derived the sinusoidal variation of Stokes $Q$ and $U$ for
each of the spectral channels. For the channels with sufficiently
strong signal, 
we derive the maximum of the $Q$ and $U$ sine curves by least--squares
fitting, and we plot them against each other. Fig.~\ref{f:TX} (bottom) shows
that the $U$ amplitudes are systematically lower than the the $Q$
amplitudes by a factor $1.20\pm0.01$. This effect is constant during the five
hours of the experiment, and we interpret it as a loss of $U$ power
due to decorrelation. The loss corresponds to an rms phase noise of 33
deg.

\subsubsection{Using grid G5}
\label{sss:G5}

In subscan 4 of the phase calibration measurement
(sect.~\ref{sss:method}),  the grid G5 is 
inserted into the beam. The cold load therefore radiates linearly
polarized power into the receivers. Its $Q$ and $U$ components are
determined from their power difference (eqn.~2) and from their
correlation (eqn.~3), while total power is simply the sum of the
powers (eqn.~1). For this fully linearly polarized signal, Stokes $I$
and $Q$ which are not affected by phase noise, can be used to predict
$U = \sqrt{I^2 - Q^2}$. The difference 
between the measured and the predicted $U$ is the decorrelation
loss. From many measurements at 86.2 GHz we obtain a decorrelation loss
of 14 \%, in agreement with the loss derived from the $\chi$ Cyg observation.

As a consistency check, we also calculate the Nasmyth (coordinate
system  $\mathcal{K_N}$) polarization angle 
\begin{equation}
\tau = \frac{1}{2}\, {\arctan}\left(\frac{U}{Q}\right)
\end{equation}
using the measured $Q$ and
the loss-corrected $U$. We obtain $54.4\dgr$ in agreement with the
measurement on grid G4 (sect.~\ref{ss:phasecal}). This good agreement
verifies the implicit assumption that the circularly polarized power
radiated by the polarized cold load is negligible ($< 1$\%).

\subsection{The signs of the Stokes parameters}
\label{ss:calPol}

The last peculiarity of polarization measurements is that contrary to
Stokes $I$ which is always positive, the other Stokes parameters can
have any sign. While the determination of the sign of $Q$ (= H $-$
V; eqn.2) is easily made by simply disconnecting one of the receivers,
the calibration of the $U$ and $V$ signs is somewhat more involved.

\subsubsection{Stokes U}
Our preferred method for determining the sign of $U$ is based on
our knowledge of the sign of $Q$  and knowing how to rotate an angle $\tau$ 
measured in the righthanded system $\mathcal{K_N}$ to the polarization 
angle $\chi$ measured in the lefthanded ($\alpha,\delta$) system 
$\mathcal{K_{\ast}}$ centered on a celestial target. The
transformation involves an even number of reflections, one change of
handedness, the Nasmyth rotation (anti--clockwise) about the elevation
angle $\epsilon$, and a (clockwise) rotation by the parallactic angle
$\eta$. This results in the relation
\begin{equation}
\tau = 90\dgr\ + \chi + \epsilon -\eta 
\end{equation} 
The relation has been verified by observing linearly polarized sources
of known position angle $\chi$, like the Crab Nebula \citep{WF} and the
moon \citep{IFpol}. The Crab nebula is our primary angle (or $U$) calibrator, 
since it is easily detected at the 30m telescope, it is strongly
polarized, and its angle
is well known ($\chi = 155$\dgr) and rather constant across the face
of the nebula \citep{WF}. An observation of the Crab Nebula is included
in every polarimetry session at the 30m telescope, and it serves as a
very powerful check of the stability of the XPOL data aquisition. 

\subsubsection{Stokes V}
For Stokes $V$ we adopt the IAU convention \citep{IAU}: $V = RHC - LHC$
where RHC (LHC) designate right (left) hand circular polarization. 
As usual, we take the electric vector of a RHC wave propagating toward the
observer to describe a counterclockwise rotation. We determined the sign
of $V$ from 3 largely independent methods: {\it (i)} using first
principles and checking them against the established usage of receivers in
VLBI observations, {\it (ii)} converting unpolarized emission from
a planet to circular of known sense with the help of a wire grid/quarterwave
plate assembly, and {\it (iii)} converting the linearly polarized
flux of the Crab nebula to circular of known sense with a quarterwave plate.
These methods, described in detail in a separate note \citep{stokesV}, 
give the same result, and the XPOL calibration software now gives the sign of
$V$ in agreement with the IAU convention.

\section{Sensitivity}

\begin{figure}
\resizebox{0.450\textwidth}{!}{\includegraphics{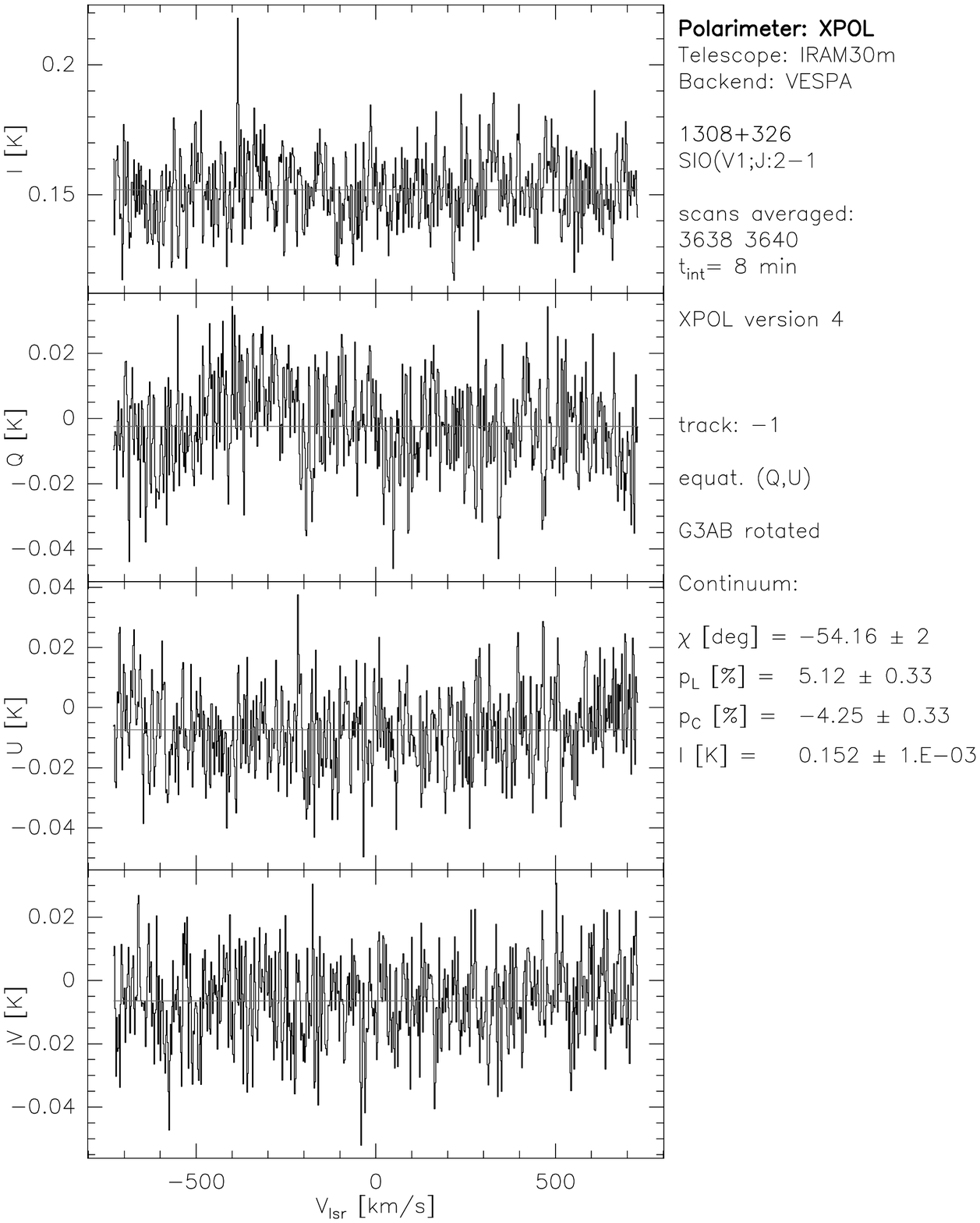}}
\resizebox{0.305\textwidth}{!}{\includegraphics{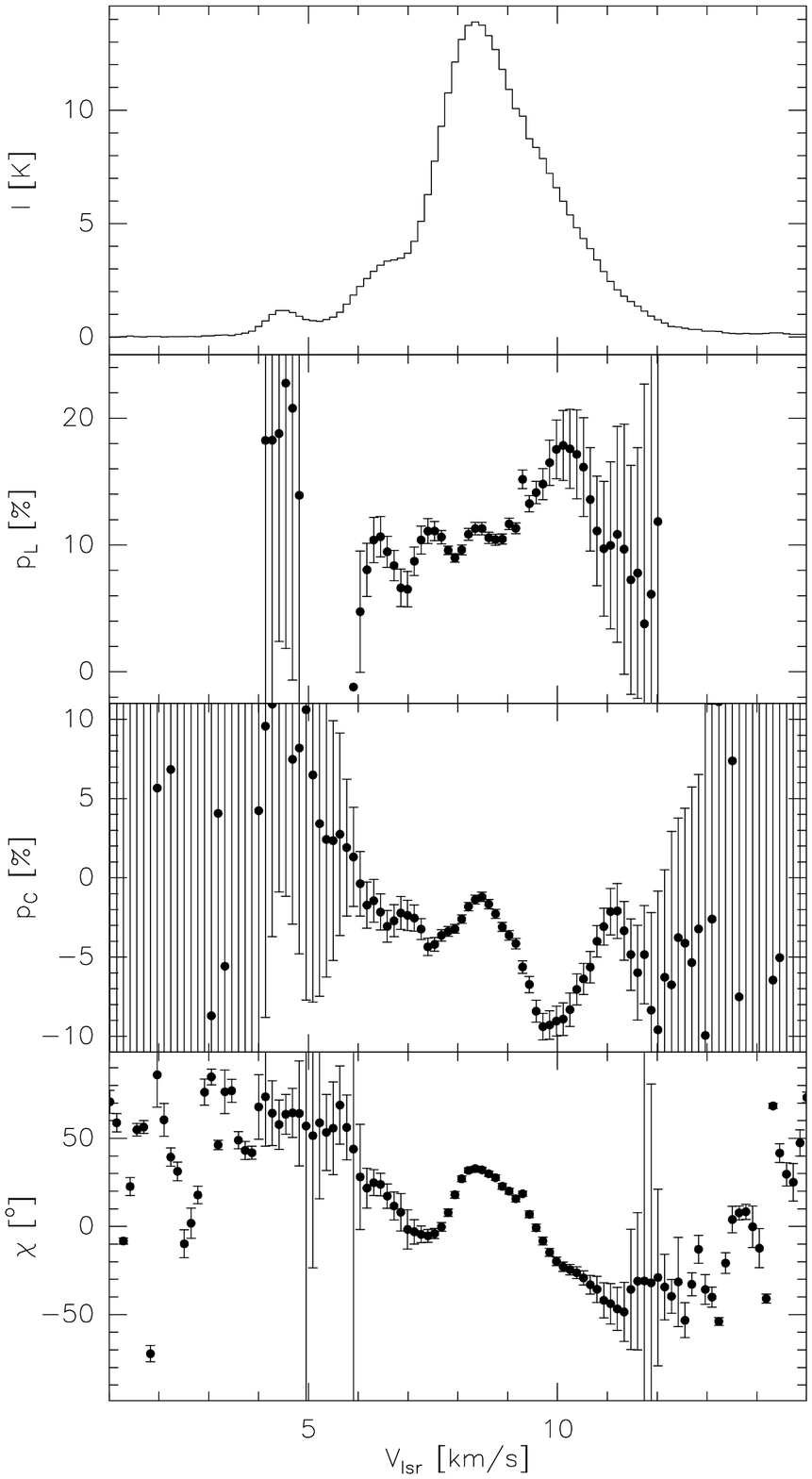}}
\caption{XPOL observations of a continuum (left) and spectral line source
 (right). The weakly polarized AGN 1308+326 was observed for 8 min.
 Its polarization parameters derived from averaging the spectral
 channels are listed on the right of the figure. 
 The SiO star V~Cam was observed with a spectral resolution of 80 kHz
 in the $v=2$, J$ =2\rightarrow1$ maser transition at 86.243 GHz. The
 linear and circular polarization degrees, \pL\ and \pC, and the
 polarization angle $\chi$\ are well defined only in the spectral
 channels where the source is bright. Error bars are large when the
 total power is weak. Integration time 4 min.
\label{f:Cobs}
}
\end{figure}

A polarimetric observation consists of 
a calibration of the antenna temperature and the phase, followed by a
wobbler--switched observation of a target. A subsequent control
measurement of the phase was soon abandoned because of the high
stability of the phase in 
the absence of large antenna motions or frequency
changes. Fig.~\ref{f:Cobs} obtained with the XPOL calibration package
shows a typical observation of an Active Galactic Nucleus (AGN) from
the list of the telescope's pointing sources ({\it left}) and a
spectral line observation ({\it right}). 
\placefigure{f:Cobs}
For the continuum observation, we used VESPA with a  bandwidth of 480
 MHz, close to the maximum IF bandwidth of the 3mm receivers.
 At this bandwidth, the channel spacing of 625 kHz
generates $\sim700$ spectral channels. The XPOL software
removes the instrumental phase from the phase of the target, derives
the calibrated $U$ and $V$ spectra, and combines them with $I$ and $Q$
to give the spectra of \pL, \pC, and the angle $\chi$. Their mean
values obtained from averaging the spectral channels are given on the
margin of the figure together with header information.\footnote{\ At the
observing frequency of 86 GHz the ratio S$_\nu$/T$_A^* = 6.0$.}

Fig.~\ref{f:Cobs} (left) illustrates the sensitivity obtainable in 3mm
continuum observations. From the radiometer formula we expect an rms
noise of 13 mK in 8 min of integration for one polarization (one
receiver), given a system temperature of  
200~K,   a spectral channel spacing of 625 kHz, wobbler switching,
and a backend efficiency of 87\%. Since each Stokes spectrum combines
the noise from two receivers, 
the noise of the Stokes spectra is expected to be 
$\sqrt{2}$ larger, in agreement with our observation where we
typically have an rms noise of 18 mK.  
The standard deviations of \pL\ and \pC\ are then derived for each
spectral channel from error propagation. In continuum observations,
the standard deviations are divided by the  
square root  of the number of channels. We thus obtain a 0.33\% formal
rms error of the polarization degree and an rms uncertainty of
1.2\dgr\ for the polarization angle. 

The observation of 1308+326 discussed here was made during good
weather with an optimized system in the context of
a large 3mm polarization survey of AGNs (Thum et al., in
preparation). The extent to which such data 
can be trusted depends on a good understanding of the systematic
errors described in the next section.

\section{Instrumental polarization}
\label{s:IP}
%

The many systematic effects which affect polarization observations are
often collectively labelled instrumental polarization. In a
rotationally symmetric antenna like the 30m telescope where ideally no
instrumental polarization is expected (sect.~\ref{ss:ideal}), such
systematic effects may occur due to subtle departures from symmetry
or measurement problems. With the XPOL setup 
where correlation polarimetry directly measures the Stokes
parameters, instumental effects are conveniently classified according
to the Stokes parameter affected. In the absence of calibration
errors, power is not destroyed or created, but only exchanged between
Stokes spectra.  The most serious problems occur when total power
is transported into other, usually much weaker Stokes parameters
(sect.~\ref{ss:feed}, sect.~\ref{ss:onaxis} and \ref{ss:beams}). Phase
errors which 
exchange  $U$ and $V$ power are negligible with XPOL due to its high
level of phase stability (sect.~\ref{sss:precision}). A
subtle effect which may transport power between $Q$ and $U$ is described in
sect.~\ref{ss:ge}. 

\subsection{The perfect Cassegrain Telescope}
\label{ss:ideal}
For axisymmetrical telescopes such as center-fed paraboloids or Cassegrain
telescopes no significant instrumental polarization can be induced by the
optical system when suitably illuminated.  This was shown for
illumination by a Huygens source\footnote{   
In this context a "Huygens source" is one for which the {\bf E} and 
{\bf H} fields are related as for a plane wave in free space.}
 in the optics approximation by
\citet{Safak} and \citet{Hanfling}  and when diffraction is
considered by \citet{MT} and \citet{Thomas}.
In the latter case a balanced HE11 mode corrugated horn was
shown to be an effective Huygens source with cross polarized
sidelobes less than $10^{-5}$ in amplitude for large telescopes ($\ge
1000$ wavelengths) where any edge currents have negligible effect 
(see \citet{Ng} for a discussion of edge effects).
The electric field in the
telescope aperture plane is then everywhere parallel and no
cross--polarization can thus exist in the far field no matter what
asymmetries may exist in the illumination or phase distribution.\ Thus
the far field 
polarization will be determined by the polarization of the effective
field distribution which illuminates the subreflector from the
secondary focus. 

\subsection{Feed imperfections}
\label{ss:feed}
The feed horn is the receiver component first encountered by the
incoming radiation. Imperfections of feedhorns may give rise to 
cross--polarization which then propagates down the rest of the signal chain.
The problem which may easily reach the 10\% level
is well known in radio astronomy \citep{CK}. Even in modern radio
interferometers, cross--polarization arising from the feeds (the
``D--terms'' of the interferometer response) is typically at the level
a few percent \citep{Cotton}.
\citet{Turlo} describe the linear
polarization characteristics of a 6cm receiver at the Effelsberg 
100m telescope.
 
Cross--polarization in the 30m receivers may be expected to be less serious
than in these cases.  For example, in XPOL, the two linear
polarizations are detected by separate receivers, each 
employing a circular corrugated horn followed by a rectangular
mono--mode waveguide which suppresses the cross polarized
field. An even stronger reduction of the cross--polar component
occurs at the grid G3
(Fig.~\ref{f:schema}). We tried to measure the residual 
cross--polarization by introducing additional wire grids right in front of the
entrance windows of the receiver dewars. No signal was detected when
looking at the hot/cold loads with the wire grids orthogonal to the
nominal polarizations of the receivers. We derive a relatively
imprecise upper limits of $\leq3$\% for the on--axis feed 
cross--polarization. More sensitive astronomical measurements are
described below.

\subsection{On--axis polarization}
\label{ss:onaxis}
\begin{figure}
\epsscale{.74}
\plotone{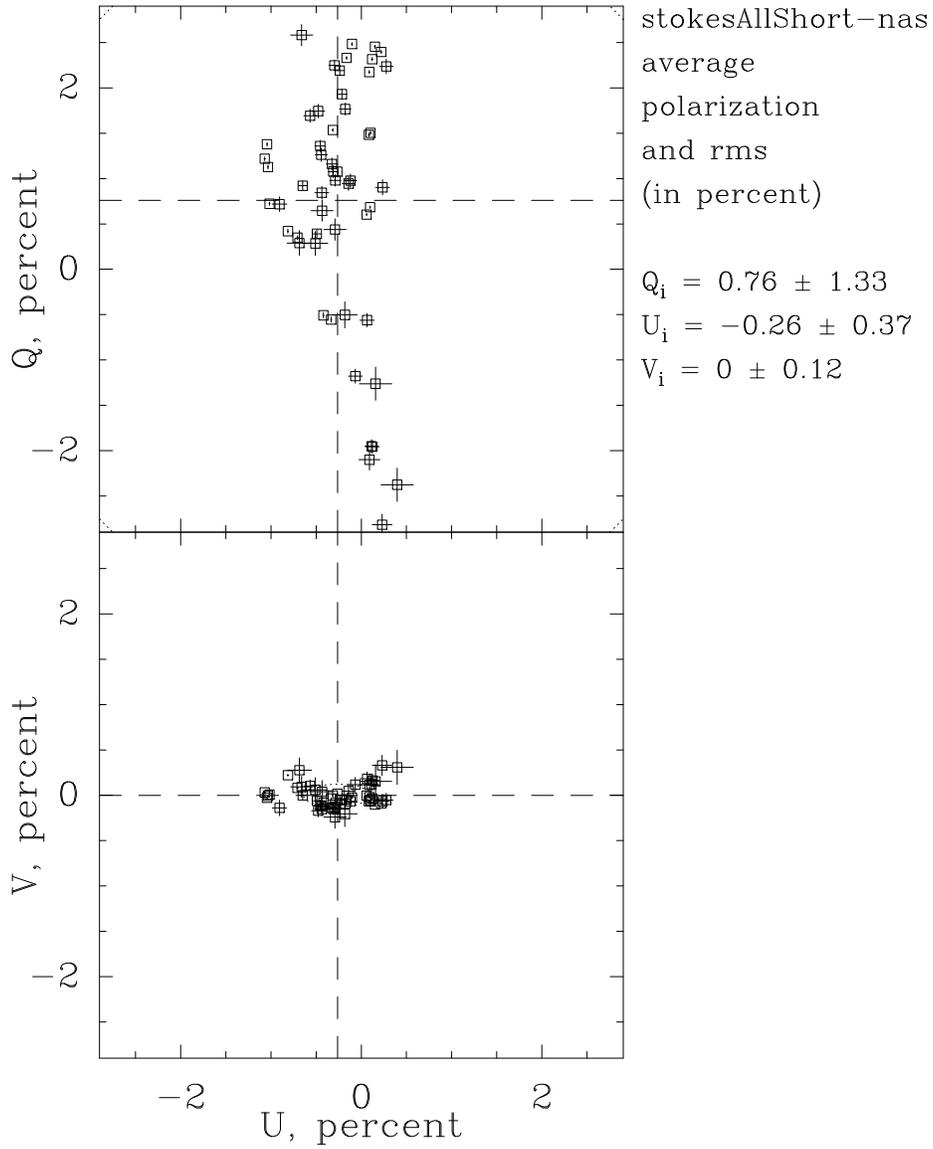}
\caption{On--axis instrumental polarization as measured from Planets
 and HII regions. Observations were made during 2 weeks in July 2005
 which included periods of very unstable and cloudy weather. All data
 are shown. The dashed lines indicate the mean values for the
 Stokes parameters $Q$, $U$, and $V$. 
\label{f:ip}
}
\end{figure}

In a  more precise study of the instrumental polarization, we looked
at the false polarization signal from unpolarized
and strong sources. In the course of the AGN survey, many such
observations were 
made on planets and HII--regions. Fig.~\ref{f:ip} shows the values 
obtained, their mean values and variances are given in Table~\ref{t:ip}. 
The data represent the on--axis instrumental polarization which is
relevant for observation of point sources.

\placetable{t:ip}

The values and their variances are drastically different for these 3
(non--$I$) Stokes parameters. We note that this pattern stayed
approximately constant during the 2 weeks of the AGN survey, and it
was also found to be the same at subsequent polarimetry sessions. The
large variance of $Q_i$ is clearly 
due to temporal variations of the atmospheric emission. This was
verified by omitting
all observations made under
unstable conditions, including many made through cumulus
clouds. Retaining only the observations obtained under stable
conditions, the variance of $Q_i$ approaches that of $U_i$, but always
remains some 10 to 30\% higher. We attribute this persistent
difference to our way of measuring 
the two parameters: $Q$ is derived from power measurements,
whereas $U$ is derived from their correlation. Fluctuations of receiver
gain or atmospheric emission which are faster than the wobbler
frequency are therefore suppressed in $U$, but not in $Q$. 

By far the best performance is obtained in $V$, where the variance
approaches the scatter expected from system noise, and the mean
instrumental offset $V_i$ is negligibly small. Circular polarization of
point sources can therefore be observed without any corrections for
instrumental polarization. This is even true for extended sources 
after the improvements described in  sect.~\ref{ss:beams}.


Despite Stokes $V$ and $U$ being both derived from correlations, their
variances are clearly  not the same. We attribute the higher variance
in $U$  to small intrinsic linear polarization of the supposedly
unpolarized sources. A few sources with \pL\simlt1\%, possibly among
the compact HII--regions in this sample \citep{Glenn}, are sufficient
to increase the sample variance to the observed value. 

Observations of linear polarization  need a stable atmosphere,
and the instrumental  polarization must be subtracted from the observed Stokes
$Q$ and $U$ values. These systematic effects set in at \pL$\simlt1.5$\%
and must be carefully checked.

\subsection{Beam maps}
\label{ss:beams}
In an effort to understand the origin of the instrumental polarization
observed on--axis (sect.~\ref{ss:onaxis}), we have made beam maps in
the four Stokes parameters. These maps were observed in the on--the--fly mode
without any switching. Reference fields were observed before each 
subscan in order to calibrate the receiver gain. The brightest
planets  which had diameters smaller than the main beam were
mapped. Fig.~\ref{f:beams} shows the 
$Q$, $U$ and $V$ maps obtained at two epochs: in 1999 when our polarimetry
tests started (left),  and in 2005 (right) obtained in the context of
the AGN survey mentioned in the previous section after the changes
described below were made.

\placefigure{f:beams}

The 1999 $Q$ and $V$ maps are characterised by strong bipolar patterns
reaching amplitudes larger than $\pm2$\%. The patterns are mostly
antisymmetric about the $y$--direction which is the vertical in the
Nasmyth cabin. Maps taken at different elevation and azimuth, and
therefore different relative orientations of the systems
$\mathcal{K_\ast}$ and 
$\mathcal{K_N}$, invariably confirm this antisymmetry about
the $y$--direction. The bulk of the polarized sidelobe power therefore
arises in components stationary in the Nasmyth cabin. 

The intersection of the $x$ and $y$--arrows is the pointing
direction of the telescope. The $Q$ map therefore suggests that the
two receivers were misaligned, mainly along the horizontal axis. The
amplitudes of the positive and negative lobes correspond to an angular 
offset of $\sim1.0''$ which is the typical precision to which receivers
can normally be aligned. 

Cross--talk between the IF signals was also investigated as a possible
cause of the instrumental polarization. Injecting a strong signal at
various points of the IF chain of one receiver, the response was
measured in the IF of the other receiver and found to be better than
$-40$ dB.

In 2005 we succeded in aligning receivers A\,100 and B\,100 using a
more accurate 
method which minimized the instantaneous difference signal from the
receivers. A precision of the alignment of $0.3''$ or better was
reached; this is  difficult to improve upon with heavy receivers mounted in
mechanically independent dewars (Nasmyth image scale: $0.7''$/mm). 

Another improvement was the orientation of the wire grid G3 
(Fig.~\ref{f:schema}). In the default setting of the 30m Nasmyth optics,
this grid must have  its wires parallel to the plane of incidence.
When a grid is used in a divergent beam (like in the case of G3),
this orientation is however not optimum with respect to cross
polarization. \citet{Chu} 
have shown that in this case the wires should be perpendicular to the plane of
incidence if enhanced cross polarized sidelobes are to be avoided (about
-24 dB in their case).
In the appendix (sect.~\ref{s:G3}) we describe in detail how
imperfections of the grid 
and its orientation introduce cross polarization, and we demonstrate
quantitatively that the polarization pattern observed with 
well--aligned receivers is fully explained by the wrongly oriented G3. 
Most of the data of the AGN survey, however,  were obtained with the
grid G3 in its optimum position.

The 2005 maps show the combined effects of the optimum alignment and
the correctly oriented G3. Striking improvements were obtained in $Q$
and $V$, and a general trend from mostly bipolar to mostly quadrupolar
is observed in all 3 maps.  With the possible exception of $Q$, the
beam patterns are no longer dominated by misalignment. This quadrupolar
polarization pattern is exactly what is  expected in a Cassegrain
antenna \citep{Olver} whose  optical layout is symmetric about the radio axis 
and which is illuminated by a feed with a $m=2$ polarization component
(see Appendix~\ref{ss:G3div}). The polarization sidelobes
observed in 2005 are thus likely to come mostly from
non--circularity of the feeds' electrical characteristics. 

The values measured at the intersection of the $x$ and $y$--axis, i.e.
the pointing direction of the telescope, are
the on--axis instrumental polarizations $Q_i$, $U_i$, and $V_i$
discussed above (sect.~\ref{ss:onaxis}). Although the observations of
the unpolarized point sources (Tab.~\ref{t:ip}) and  the
maps were not made  contemporaneously, the agreement between the two sets
of values for the instrumental polarization is
remarkably good. In the center of the $Q$ map, there is a large slope
where angle--of--arrival fluctuations can induce random
$Q_i$ values. This is not the case for $U$. This difference may help to explain
why we find the variance of $Q_i$  to be always larger than that in
$U_i$, even under good atmospheric conditions.
 

\subsection{Gain variation with elevation}
\label{ss:ge}
The IRAM 30m telescope has a homologous design which is optimized for
an elevation of 43\dgr. The residual surface errors introduce a
degradation of the on--axis gain of the antenna, or its aperture
efficiency $\eta_a$, when moving away from the design elevation. 
\citet{ge} showed that at $\lambda3$mm $\eta_a$ decreases by about 4 --
5\% at the extreme elevations relative to its optimum value near the design
elevation.  

Given this marked dependence of $\eta_a$ on elevation, we
suspect that the residual deformations of the structure may be
slightly asymmetric  in the vertical and horizontal directions. This would give
rise to instrumental polarization, primarily in $Q$ when observed in
the horizontal system. The receivers see this polarization after the Nasmyth
rotation as a mixture of $Q$ and $U$ which has a well--defined
elevation dependence. We have therefore re--analysed the instrumental
polarization data presented in sect.~\ref{ss:onaxis}, but we do not
find any dependence on elevation stronger than 1.5\%. The precision of
this upper limit is however strongly limited by the heterogeneity of
the data and by the incomplete coverage of the elevation range.

A new opportunity for a more precise measurement of this effect
occured in May 2005 when the quasar 3C454.3 had an
outburst. We made 45 dedicated XPOL measurements over the elevation
range from 10\dgr\ to 69\dgr. Due to the brightness of the source
($S_\nu \simeq 15$ Jy) the precision of an individual polarization
measurement was $\la0.1$\%, and the linear polarization of the quasar
(\pL\ =4\%, $\chi=60$\degr) was easily detected. 
The suspected elevation dependence of the instrumental polarization
would show up as an elevation modulation of \pL\ around its mean
value. Indeed such
modulations were found at levels of 1.5\% ($Q$) and 0.5\%
($U$). Unfortunately, these modulations are indistinguishable from
short--timescale intrinsic polarization variations of
the quasar. Indeed, our measurements show that the total power (Stokes
$I$) of the quasar varied over the 12 hours monitoring period by more
than 20\%. We therefore take these quasar measurements only to confirm
our previous upper limit of 1.5\% for any polarization--dependent gain
variation with elevation.

\section{Conclusions}

XPOL, the first correlation polarimeter on a large single dish millimeter
telescope, is a versatile instrument, capable of measuring line and
continuum in all four Stokes parameters simultaneously. The setup makes
use of the Observatory's 3mm (single pixel) heterodyne receivers which
are housed in separate dewars on mechanically independent mounts. Our
study shows that these receivers and their arrangement, which were not
designed with polarization observations in mind, can nevertheless be
used efficiently for many types of polarization observations.

In particular, all issues related to calibration were solved
satisfactorily. The instrumental phase is precisely measured ($\simlt1$\dgr)
for all spectral channels individually, and it is very stable in time.
A new method is applied for precisely measuring the decorrelation
losses.

The many facets of instrumental polarization, in practice the factors
limiting the precision of XPOL observations, were studied at 3mm
wavelength for all Stokes parameters in much detail. The most
information can be drawn from maps of the Stokes beams. We showed that
the pattern and strength of the polarization sidelobes very
sensitively depends on the precision of the alignment between the
receivers. Only when a precision of $\simlt0.3''$ was reached, did we
see the 4--lobed pattern expected in a Cassegrain telescope with its
symmetry about the radio axis. This symmetry does not appear to be destroyed by
additional reflections on the flat mirrors of our Nasmyth focus.
The 4--lobed structure of
the polarization sidelobes and their maxima are then determined by
slight differences in the electric parameter (polarization, beam
width, etc.) of the feedhorns.  

The practical limits for XPOL observations of point sources can be
inferred from the values in Stokes $Q,\ U$ and $V$ at the center of the 
Stokes maps. These values are compatible with the ``false
polarization'' measured on strong and unpolarized point sources. We
conclude that Stokes $Q$, which is measured as a power difference, is
most affected by instrumental polarization, whereas Stokes $U$ and $V$,
which are measured as  cross correlations, are very little affected. The
lower instrumental polarization of $U$ and $V$ is due to the insensitivity  of
correlations to fluctuations of receiver gain and atmospheric
emission. An additional disadvantage arises for $Q$ under an unstable
atmosphere where the wobbling period may be too long for complete
cancellation of the emission fluctuations.

In summary, XPOL observations of Stokes $U$ and $V$ are possible down
to the $\sim0.5$\% level before systematic effects set in, both for
point--like and extended sources. Instrumental polarization limits
Stokes $Q$ observations at the 1.5\% level. With proper
precaution, this limitation can be overcome for point
sources. Measurements of the linear polarization of extended sources
are the most difficult observations with XPOL.

The design of a new receiver
optimized for polarization  would have to pay special attention to the
feed, as its cross polarization constitutes the ultimate limit for
systematic measurement errors. The alignment of the two feed horns,
the next largest source of systematic error, appears easier if the
feed horns (and the subsequent radiofrequency components) are housed
in the same dewar, or even better, only one feed is used for the two
polarizations. It therefore appears quite feasible on a big single dish
millimeter telescope for a carefully designed polarization receiver
to keep instrumental polarization well below the 0.5\% level.

\acknowledgements
We wish to acknowledge the unfailing support by Manuel Ruiz and his team
of telescope operators during the long and sometimes strenuous
commissioning of XPOL. M. Torres (IRAM) helped through frequent discussions
and by building a phase shifter used in an initial analog
version of XPOL. The support by the IRAM backend group was essential
in the construction of the VESPA backend, including its capability of
making cross correlations. We also thank M. Carter (IRAM) for
discussions and help.


\appendix

\section{Instrumental polarization generated in a wire grid}
\label{s:G3}

In the optical setup of the XPOL polarimeter, the 
wire grid G3 (Fig.~\ref{f:schema}) has an important function, since it
is used to split the incoming radiation into its  vertical and
horizontal components which are subsequently 
recorded by receivers A\,100 and B\,100.  Any imperfections of G3 
generate cross polarization which may be recorded as instrumental
polarization. We therefore studied the polarization performance of
wire grids in some detail, both when illuminated by parallel beams
(sect.~\ref{ss:G3para}) and by divergent beams (sect.~\ref{ss:G3div}).

\subsection{Grid in a parallel beam}
\label{ss:G3para}

 The bandwidth of a wire grid is
 potentially very large,\ the upper frequency of operation being
 determined by the intrinsic accuracy with which the grid can be made,
 and perhaps by the limited conductivity of the wires.\ However at
 millimeter wavelengths, we are close to these limits, and 
 the splitter grid G3 and also the other grids in the receiving
 system,  have only a finite discrimination against the unwanted
 linear polarization.\ There is first of all the intrinsic
 discrimination for a perfect grid,\ and secondly that due to errors
 in manufacture which render the grids non-ideal.
 Both effects have been studied by \citet{Houde}.

One polarization is selected by reflection,\ when the discrimination
is measured by the reflection coefficient for polarization perpendicular to
the wires $R_{\perp}$.\ The orthogonal polarization is selected by
transmission when the transmission coefficient $T_{\parallel}$ for parallel
polarization is appropriate.\ From the equations 59,\ 60,\ 66 and 67
of \citet{Houde} the amplitude reflection (R) and transmission (T)
coefficients are for  plane wave illumination
 
\begin{eqnarray}
   R_{\perp}     & = &  -i(1-\alpha^2)x/\gamma		\label{A1}\\	
   R_{\parallel} & = &  -(1-i2\gamma Dz)/(1+(2\gamma Dz)^2)\label{A2}\\	
   T_{\parallel} & = &  1+R_{\parallel}	     \label{A3}		\\
   T_{\perp}     & = &  1-R_{\perp}   \label{A4}
\end{eqnarray}
where for the splitter,
\ $\gamma = 1/ \sqrt{2}$,\ the wire's radius is $a$,\ their separation
is $d$,\ and $x= \pi ^2 a^2/ {\lambda d}$ , 
\ $z=log_e({{d} \over{2 \pi a}})$ and $D={d \over \lambda}$.  For the
(favorable) case where the grid is rotated 45\dgr\ around an axis parallel
to the wires $\alpha={1 \over \sqrt{2}}$ otherwise $\alpha=0$.
 
 The response of a complex cross correlator to unpolarized radiation
 with components $e_{\parallel},\ e_{\perp}$ is
\begin{eqnarray}
    \Gamma  & = & e_{\parallel}^2 R_{\parallel}T_{\parallel}^{*}
                  + e_{\perp}^2 R_{\perp}T_{\perp}^{*}   \nonumber
\end{eqnarray} 
which can be expressed as components of the Mueller matrix
\citep{Tinbergen, Heiles-Mueller} as
\begin{eqnarray}
     M_{IU}+iM_{IV} & = & {I \over 2}\big[R_{\parallel}
     (1+R_{\parallel}^{*})+R_{\perp}(1-R_{\perp}^{*})\big]  \nonumber	
\end{eqnarray} 
For $d=0.1$mm,\ $a=0.0125$\,mm and frequency 86 GHz, such as used with
XPOL, eqs.~\ref{A1}--\ref{A4} yield 
$ R_{\perp} = -ix = -i\ 0.0034 ,\ R_{\parallel} = -1 +i\ 0.00976,\
T_{\perp} = 1 + i\ 0.0034,\ T_{\parallel} =  -i\ 0.00976$ 
and thus
\begin{eqnarray}
{M_{IU}+iM_{IV} \over I} =0.000053+i\ 0.0032		\label{A5}  
\end{eqnarray}
 The effects of random errors in the spacing of the wires may lead
to larger values for $T_{\parallel}$ (and perhaps $R_{\perp}$ also).
\citet{Houde} give some representative values for transmittance
$T_{\parallel}^{2}$ in their 
Figure 3,\ ($a=0.0125$mm). An approximate interpolation and scaling to the
frequency of the XPOL splitter G3 give $T_{\parallel}^{2} = 1.8 \%$ for
a $5 \%$ error in spacing,\ and $T_{\parallel}^{2} = 3.2 \%$ for a $16
\%$ error.  Thus the polarization splitter grid should give a polarization
 discrimination of at best $0.3 \%$ (eq.~\ref{A5}). More realistically
 it is perhaps $2-3\%$, when up to $16\%$ construction errors are considered.

The H and V optical branches between the polarization grid and
the  correlator will however provide further polarization
discrimination due to  additional grids and the receiver horns. Even
if this additional rejection  amounts to only $5 \%$ cross
polarization, the overall cross  polarization expected for the system
would be at most $0.1  \%$.  So the initially measured values of the
order of up to $2 \%$ cross  polarized sidelobes are unexpected.

\subsection{Grids in a divergent beam}
\label{ss:G3div}

\begin{figure}
\epsscale{.71}
\plotone{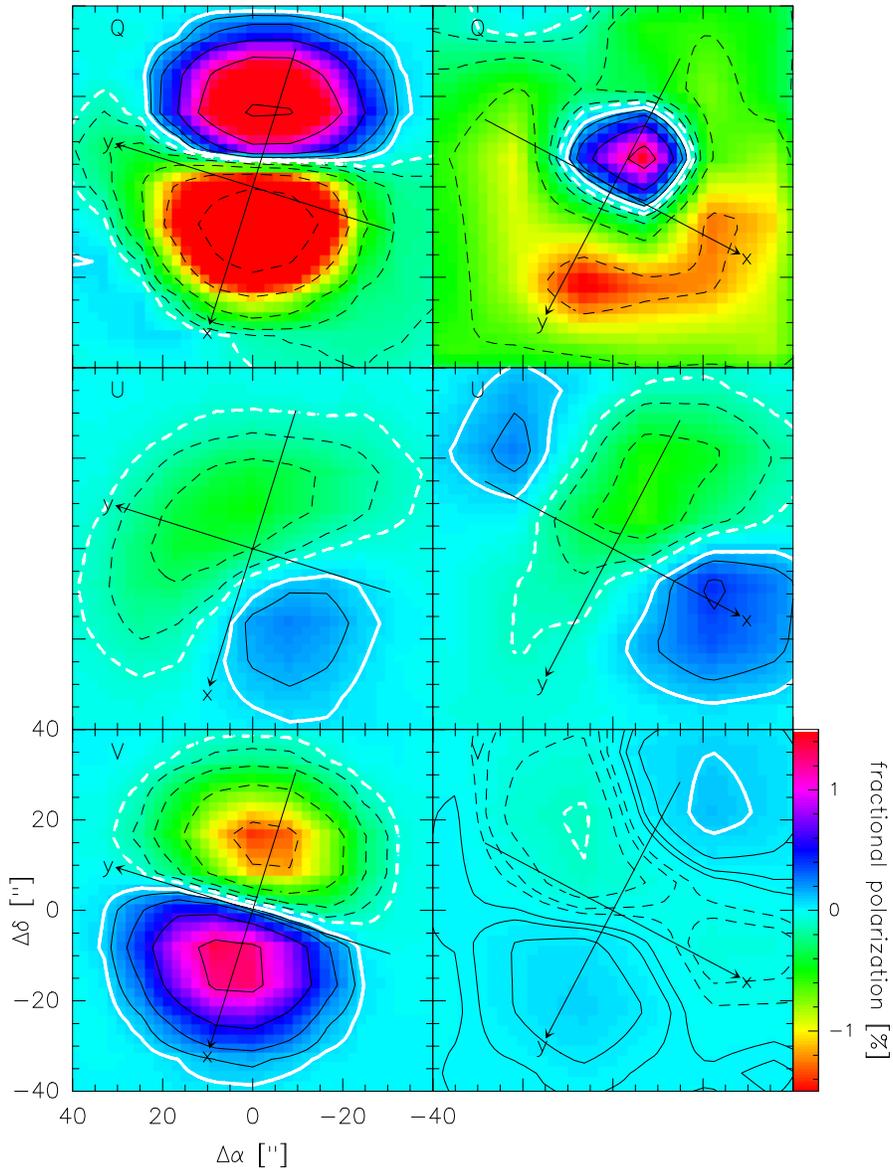}
\caption{Beam maps of the Stokes parameters $Q$, $U$, and $V$
 (from top to bottom)
 obtained at 86 GHz in 1999 (left) and 2005 (right). All maps have the
 same flux scale indicated on the lower right. Contours are drawn at
 0.2, 0.4, 0.8 1.6, and 3.2\%, negative contours are dashed. The thick
 white contours are at $\pm0.1$\%. The orientation of the
 $x$ and $y$ axes of the Nasmyth coordinate system $\mathcal{K_N}$ is shown.
 Causes of the better results obtained in 2005 are described in the
 text (sect.~\ref{ss:beams}). 
\label{f:beams}
}
\end{figure}

 Grids such as that used in the polarization splitter are very good at 
"cleaning up" the linear polarization of a parallel beam. However in a 
divergent beam, as in the Pico Veleta receiving system, they can give rise 
to quite significant cross polarization. This was shown both experimentally,
and theoretically, by \citet{Chu}.  They show (see their eqs.~9 and
10) that the electric fields parallel and perpendicular to the wires
are as follows:
\begin{eqnarray}
      E_{\parallel} &=& -E_{in}cos(\beta)\big[1-sin^{2}(\phi)(1-cos(\theta)
                       -sin(\theta)sin(\phi)cot(\gamma)\big] \label{A6}\\
      E_{\perp} & =& -E_{in}cos(\beta)\big[sin(\phi)cos(\phi)(1-cos(\theta))
                       +sin(\theta)sin(\phi)cot(\gamma)\big] \label{A7}
\end{eqnarray} 
Here $\beta$ is the inclination of the grid to the beam axis (45\dgr),
\ $\theta$ and $\phi$ define directions relative the beam axis, and $\gamma$
specifies the orientation of the wires in the plane of the grid. If the
wires are perpendicular to the incident beam axis (perpendicular to the
plane of incidence), then $cot(\gamma)=0$ and only a minimal
second--order cross--polarization is present.  However for the alternative
orientation of the wires $cot(\gamma)=1$,\ a large first--order 
cross--polarization is present after the beam reflects from the grid.
The first--order cross--polarization is proportional to the sin of the 
angle $\theta$ relative to the beam axis,\ and has a $m=1$ variation 
around the beam axis ($cos(\phi)$).
This dipolar ($m=1$) symmetry agrees with that observed for $M_{IU}$
and $M_{IV}$ (Fig.~\ref{f:beams}, left column). 
Furthermore, since at the 30m telescope $\theta$ reaches about 3\dgr,
it seems possible to generate a few percent of  instrumental polarization 
by this mechanism.
 
We estimate the magnitude of the cross--polarization from a model of the
electric field distributions $\bf{E}_A(x,y),\ \bf{E}_B(x,y)$ in the 
telescope aperture associated with receivers A\,100 and B\,100.
The fields are calculated using equations \ref{A6} and \ref{A7}.
The effects of differential
pointing errors and defocus can be simulated by the addition of linear 
or quadratic phase terms.  Identical Gaussian tapers have been assumed, 
but of course other tapers and field distributions can be accommodated. 
The corresponding far field radiation patterns calculated by Fourier 
Transform are then $\bf{A} (\theta, \phi)$,\ and $\bf{B} (\theta, \phi)$.\ 
The inputs to the correlator, when the telescope is illuminated by 
unpolarized radiation with (uncorrelated) components $e_{\times}\ ,e_{co}$ 
from the direction $\theta,\phi$ are
\begin{center} \(
\begin{array}{rcl}
 V_a &=&  e_{\times}\ A_{\times} (\theta,\phi) + 
          e_{co}\ A_{co} (\theta,\phi)		\\
 V_b &=&  e_{\times}\ B_{co} (\theta,\phi+{\pi \over 2}) - 
          e_{co}\ B_{\times} (\theta,\phi+{\pi \over 2})	
\end{array}\) \end{center} 
from which we derive the relevant components of the corresponding Mueller
matrix as:
\begin{center} \(
\begin{array}{rcl}
 I(\theta,\phi)&=& <V_a^2> + <V_b^2>                              \\	
 M_{IQ}(\theta,\phi)&=& <V_a^2> - <V_b^2>	                  \\
 M_{IU}(\theta,\phi)+i\ M_{IV}(\theta,\phi)&=& <V_a\ V^*_b>       \\
&=&    {I \over 2}
   \big[A_{\times}(\theta,\phi)B^*_{co}(\theta,\phi+{\pi \over 2})
       -A_{co}(\theta,\phi)B^*_{\times}(\theta,\phi+{\pi \over 2})\big]
\end{array}\) \end{center}
Here we have taken the two beams to be rotated (about their axes by 90\dgr) 
to have orthogonal linear polarization.

 Any periodicity of $M_{IU}(\theta,\phi)$ or $M_{IV}(\theta,\phi)$ in 
the angle $\phi$ about their axes will be essentially determined by 
that of the cross--polarized beams $B_{\times}$,  since the co-polarized 
beams will to a first approximation be invariant with $\phi$. 
The far field beams are just rescaled and smoothed images of the 
secondary focus field distribution in amplitude and polarization,
provided that this field distribution radiates as a Huygens source. 
Then the far field will have also the same periodicity.\ That is,\ 
when expressed as series in $cos(m \phi),\ sin(m \phi)$ they will 
have the same $m$ values.\ In the present case where the effective 
source at the secondary focus is approximately that of the beam waist 
of a Gaussain beam, the assumption of a Huygens source may be justified.

 Many imperfections in the optical system leading to cross polarization 
might be expected to correspond to higher order modes of circular 
waveguides.
For example corrugated waveguides operating off the design frequency would
radiate an excess of the $TE_{11}$ or $TH_{11}$ mode,\ both of which have
an $m$ value,\ as defined above,\ of 2.\ The only modes with $m=1$ are
$TE_{01} (H_{01})$ and $TH_{01} (E_{01})$,  neither of which radiate 
along the beam axis and which are very difficult to excite.\ 
An exception is provided by an offset elliptical mirror which gives 
a field distribution with a fan like divergence of field lines 
across the beam  --- with a large $m=1$ component \citep{Murphy}.
In a similar way, the polarization splitter grid can induce such 
cross polarization when used at a non-optimum orientation 
($cot\,\gamma=1$).  A similar $m=1$ asymmetry   could result 
from a lateral displacement in the telescope aperture of a $m=2$ 
cross--polarized electric field pattern.     

\begin{figure}
\epsscale{.80}
\plotone{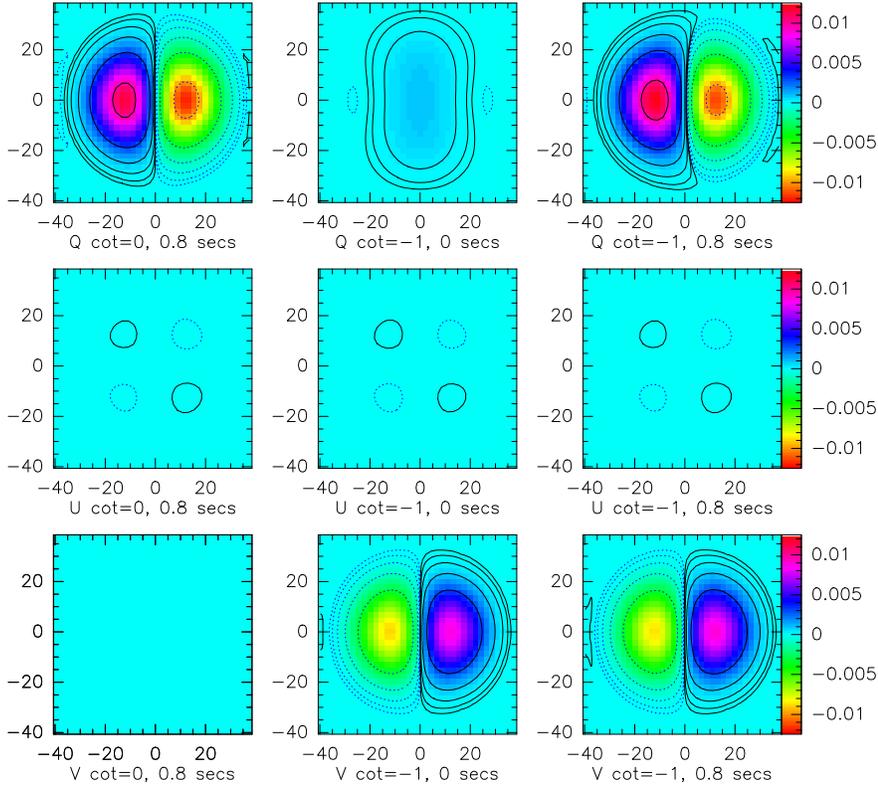}
\caption{Simulation results for the far field cross--polarized sidelobes.
  The effects of changing the orientation of the polarization grid ($cot$)
  and the pointing offsets (arc seconds) between the 2 receivers are shown
  in the 3 columns.\ Left column - offset 0.8 arc seconds,\ optimum 
  orientation,\ middle
  column,\ - no offset,\ non-optimum orientation,\ right column - offset 0.8
  arc seconds,\ non-optimum orientation.\ From top row to bottom row are
  displayed the Muller matrix elements $M_{IQ},\ M_{IU},\ M_{IV}$.\ The
  contours are at power levels relative to $I$ of $\pm 0.01 ,\ \pm 0.0036
  ,\ \pm 0.001 ,\ \pm 0.00036,\ \pm 0.0001 ,\ \pm 0.000036$.
\label{f:app}
}
\end{figure}

 Fig.~\ref{f:app} illustrates the roles of the polarization splitter
 and receiver misalignment.
It displays the results from three simple models. 
The far field radiation patterns for each receiver (A and B) were
 calculated by Fourier Transform of corresponding electric field
 distributions in the aperture plane.\ The equations of\   
 \citet{Chu} were used to describe their polarization and linear
 phase gradients were applied to simulate pointing offsets.\ A
 Gaussian illumination (14 dB taper) was assumed. 
The lefthand column of Fig.~\ref{f:app} 
has $M_{IQ},\ M_{IU},\ M_{IV}$ for the case of 0.8 arc seconds offsets 
between receivers A\,100 and B\,100, and for optimum orientation of the 
polarization splitter.  The $M_{IV}$ values are in this case zero within 
the rounding errors of calculation. The middle column has
results for non-optimum orientation and no misalignment. The righthand
 column includes in 
addition a pointing offset of 0.8 arc seconds between the two receivers.\ 
The overall similarity between the latter model maps and the maps of
Fig.~\ref{f:beams} (left column) demonstrates that both effects, an
alignment error and a wrong orientation of the splitter G3, are needed
to explain the 1999 maps.  

 Since the polarimeter is essentially a zero spacing interferometer, 
many of the above considerations will apply also to interferometers 
with spaced antennas.

\clearpage


\begin{table}
\caption{On--axis instrumental polarization, in percent. \label{t:ip}}
\begin{tabular}{lcrcr}\\[1ex]\hline
Q$_i$ & = & 0.76  & $\pm$ & 1.33 \\
U$_i$ & = & -0.26 & $\pm$ & 0.37 \\
V$_i$ & = & -0.03  & $\pm$ & 0.12\\[1ex]\hline
\end{tabular}
\end{table}

\clearpage


\clearpage


\clearpage

\clearpage


\end{document}